\documentclass[11pt]{article}
\usepackage[dvips]{graphicx}
\usepackage{amsmath,amssymb,amsfonts}
\usepackage{color}
\begin{document}

\title{Bounds on quantum gravity parameter from the $SU(2)$ NJL effective
model of QCD}
\author{K. Nozari\thanks{knozari@umz.ac.ir},\quad M.
Khodadi\thanks{m.khodadi@stu.umz.ac.ir}\quad and \quad
M. A. Gorji\thanks{m.gorji@stu.umz.ac.ir}\\
{\small {\it Department of Physics, Faculty of Basic
Sciences, University of Mazandaran,}}\\
{\small {\it P. O. Box 47416-95447, Babolsar, Iran}}}

\maketitle
\begin{abstract}
Existence of a minimal measurable length, as an effective cutoff in the ultraviolet
regime, is a common feature of all approaches to the quantum gravity proposal. It
is widely believed that this length scale will be of the order of the Planck
length $\lambda=\lambda_0\,l_{_{\rm Pl}}$, where $\lambda_0\sim{\mathcal O}(1)$ is
a dimensionless parameter that should be fixed only by the experiments. This issue
can be taken into account through the deformed momentum spaces with compact
topologies. In this paper, we consider minimum length effects on the physical
quantities related to three parameters of the $SU(2)$ Nambu-Jona-Lasinio effective
model of QCD by means of the deformed measure which is defined on compact momentum
space with ${\mathbf S}^3$ topology. This measure is suggested by the doubly special
relativity theories, Snyder deformed spaces, and the deformed algebra that is obtained
in the light of the stability theory of Lie algebras. Using the current experimental
data of the particle physics collaboration, we constraint quantum gravity parameter
$\lambda_0$ and we compare our results with bounds that are arisen from the other
experimental setups.
\vspace{5mm}\\
\begin{description}
\item[{\bf Key Words}]:
Phenomenology of Quantum Gravity, Quark Condensate, Pion Decay
Constant, Pion Mass.
\item[{\bf PACS numbers}]:
04.60.-m, 04.60.Bc, 12.39.-x, 12.40.-y
\end{description}
\end{abstract}

\section{Introduction}
Quantum gravity (QG) candidates such as string theory and loop quantum gravity strongly suggest
the existence of a minimum length scale below which no other length can be observed \cite{ST,LQG}.
It is then natural to expect that a non-gravitational theory, which includes an invariant
minimum length scale, arises at the weak gravity limit (but high energy regime) of the ultimate
QG theory. Such an effective theory will be reduced to the standard well-known theories at the
low energy regime in the light of the correspondence principle. In the absence of a full quantum
theory of gravity, one may do in reverse: Starting from quantum mechanics or special relativity
and deforming them in such a way that they include an invariant minimal length scale. The first
attempt in this direction was taken by Snyder in 1947 who formulated a discrete Lorentz-invariant
spacetime \cite{Snyder}. Quantum field theories turn out to be naturally ultraviolet-regularized
in this setup \cite{Snyder-QFT}. Motivated by the string theories, generalized uncertainty
relations are suggested that support the existence of a minimal length through the nonzero
uncertainty in position measurement \cite{GUP,GUP-HS}. The polymer quantum mechanics is
investigated in the symmetric sector of loop quantum gravity which also supports the existence
of a minimal length scale known as the polymer length scale \cite{QPR}. The relation between
the generalized uncertainty relation and the polymer quantization scenario is also shown in Ref.
\cite{PLR-GUP}. Furthermore, the doubly special relativity theories are formulated in order to
take into account a minimal observer-independent length scale in special relativity framework
\cite{DSR}. The noncommutative phase spaces are the other interesting framework to take into account a
minimal length scale \cite{NC-QM}. Apart from the details of the above mentioned models, all of
them are in agreement in the existence of a minimal measurable length. The question then arises
is: How much a minimal length would be small? The conclusive answer to this question will
became clear just after formulating a full QG theory. Nevertheless, one can constrain QG
parameter by means of the {\it correspondence principle}. More precisely, the effective
approaches are investigated by deforming the well-known theories to include a deformation
parameter that signals the QG fundamental scale. Thus, one expects that the deformation
parameter will disappear at low energy regimes since the QG effects are negligible at this
regime. In other words, the effective theories should reduce to the standard non-deformed ones
at the low energy regime and also they would not destroy the prediction of the corresponding
standard model at this regime. In this respect, one can obtain an upper bound on the QG
parameter in any well-tested low energy regime's experiment. In recent years, many attempts
have been done in this direction within the various experimental setups, see for instance Refs.
\cite{B-GUP1,B-GUP2} in which the upper bounds are found on the QG parameter in the context of
the generalized uncertainty principle (see also Refs. \cite{B-NC0,B-NC} for the case of the
noncommutative spaces). There are two determinant factors in these considerations: The energy
scale of the experiment and the accuracy of the measurement in the proposed experiment. While
the former is usually fixed for a particular experiment, the latter could be improved by
upgrading the instruments of the experimental setups. Thus, it is plausible to expect that an
experimental setup at the higher energy scales will constrain the QG parameter with more
accuracy. In this sense, we have explored an upper bound on the QG length scale in the
framework of Nambu-Jona-Lasinio (NJL) phenomenological model of quantum chromodynamics (QCD).
The NJL model is non-renormalizable when it is applied to the thermodynamics and it is then
convenient to introduce a three-momentum cutoff $\Lambda$. Evidently, the three-momentum
cutoff $\Lambda$ in NJL model cannot exceed $1$ GeV which is very small with respect to
the expected energy scale of QG ($E_{_{\rm Pl}}\sim\,10^{19}$ GeV). Although QG effects are
very small in this energy regime, we are interested to answer the question that how much they
would be small in order to respect the NJL model predictions?

The structure of the paper is as follows: In section 2, we briefly review the $SU(2)$ two
flavors NJL model. In section 3, we introduce a deformed measure that includes minimal length
effects and is suggested by some effective theories of QG. Using the deformed measure, we
find the QG corrections to the dependent parameters of the NJL model in section 4 which
allows us to constrain the QG dimensionless parameter $\lambda_{0}$. Section 5 is devoted to
the summary and conclusions\footnote{We work in the units $\hbar=1=c$.}.

\section{NJL Model}
QCD is a theoretical framework of the strong nuclear force for hadrons in which quarks
interact with non-Abelian $SU(3)$ gauge fields known as gluons. At the high energy regime,
QCD has asymptotic freedom property, i.e. the running coupling of QCD decreases at short
distances \cite{Mandel}. Therefore, the perturbation theory is applicable for the high
energy phenomena with momentum transfer $q\gg\Lambda_{QCD}$, where $\Lambda_{QCD}\approx
100-200$ MeV is a typical energy scale of QCD. QCD has two important properties at low
energy regime: The confinement and the spontaneous breaking of the chiral symmetry. The
quarks and gluons are enclosed inside the packages that are called hadrons (quark-gluon
bound states) through the confinement property. Beside, the large effective masses of the
quarks and also the light masses of the pseudoscalar mesons originate from the spontaneous
breaking of the chiral symmetry. Nevertheless, QCD is non-perturbative at low energy regime
since the strong coupling constant increases when the energy scale approaches to
$\Lambda_{QCD}$. To remedy this problem, the effective phenomenological models of QCD are
investigated. The so-called Lattice QCD (LQCD) is the most well-known candidate that is
investigated to solve the non-perturbative feature at confinement phase \cite{Wilson} (see
Ref. \cite{Gupta} for review). In particular, the dynamic generation of the fermion masses
are explained by the chiral symmetry breaking in NJL model with which we are interested in
this paper. Evidently, the behavior of the mesons in the hot and dense matters are well
understood in this setup \cite{volk}. Apart from the lack of the confining mechanism in
this model, it is an appropriate phenomenological model to study the low energy aspects of
the hadrons physics (see Refs. \cite{NJL} for details). The NJL effective phenomenological
model of QCD is originally proposed before formulating the QCD to explain interactions of
the nucleons with mesons \cite{Nambu}. Today, however, the model is defined by the
lagrangian formalism of QCD but with the important difference that now fermionic degrees
of freedom are two (or three) flavors and three colors of the quark fields
\cite{Klevansky,Hatsuda} \footnote{For the sake of simplicity, we restrict ourselves to
the two light flavors in this work.}. The simplest form of the Lagrangian for the $SU(2)$
NJL model with two quark flavors and a nonzero bare quark mass $m$ is given by \cite{NJL}
\begin{equation}\label{a-NJL}
{\mathcal L}_{_{NJL}}=\bar{\psi}(i\gamma_{\alpha}\partial^{\alpha}-m)\psi+G
\left[(\bar{\psi}\psi)^{2}+(\bar{\psi}i\gamma_{5}\vec{\tau}\psi)^{2}\right],
\end{equation}
where quark fields $\psi$ are Dirac spinors carrying colors and $\vec{\tau}$ are the
Pauli spin matrices. The linearized field equations for quark fields $\psi_{u}$ and
$\psi_{d}$ are now read as
\begin{align}\label{b-NJL}
&(i\gamma_{\alpha}\partial^{\alpha}-m_{u})\psi_{u}+2G\left[\langle
\bar{\psi}_{u}\psi_{u}\rangle+\langle\bar{\psi}_{d}\psi_{d}\rangle
\right]\psi_{u}=0, \hspace{3cm}\nonumber\\&(i\gamma_{\alpha}
\partial^{\alpha}-m_{d})\psi_{d}+2G\left[\langle\bar{\psi}_{d}
\psi_{d}\rangle+\langle\bar{\psi}_{u}\psi_{u}\rangle\right]\psi_{
d}=0,&
\end{align}
so that the dynamical quark masses are given by
\begin{align}\label{c-NJL}
&M_{u}=m_{u}-2G \left[\langle\bar{\psi}_{u}\psi_{u}\rangle+
\langle\bar{\psi}_{d}\psi_{d}\rangle\right],\hspace{3cm}\nonumber\\
&M_{d}=m_{d}-2G \left[\langle\bar{\psi}_{u}\psi_{u}\rangle+
\langle\bar{\psi}_{d}\psi_{d}\rangle\right].&
\end{align}
The numerical value of the quark masses $M_{u,d}$ are known as constituent quark masses
since they are almost equal to the common values in the non-relativistic quark models.
The NJL model defined in (\ref{a-NJL}) includes three parameters: the bare quark mass
$m$, the coupling strength (or quark-quark coupling) $G$ and the three-momentum cutoff
$\Lambda$. Notice that the bare quark masses are usually take to be $m_{u}=m_{d}=m$ to
respect the isospin symmetry. These three NJL parameters are usually fixed to reproduce
the chiral physics in the hadronic portion. The physical quantities quark (chiral)
condensate $\langle\bar{\psi}\psi \rangle$ and pion decay constant $f_{\pi}$ are used
to fix the two dependent parameters $\Lambda$ and $G$ of the NJL through the relations
\begin{equation}\label{d-NJL}
\langle\bar{\psi}\psi\rangle_0=\langle\bar{u}u\rangle_0=\langle
\bar{d}d\rangle_0=-\frac{N_cM}{\pi^2}\int_{0}^{\Lambda}\frac{dp
\,p^2}{E}\,,
\end{equation}
\begin{equation}\label{e-NJL}
f_{0\pi}^2=\frac{N_cM^2}{2\pi^2}\int_{0}^{\Lambda}\frac{dp\,p^2
}{E^3}\,,
\end{equation}
where $N_c=3$ denotes the color degrees of freedom and $E=\sqrt{p^2+M^2}$ is the energy
of a quark and anti quark of flavor $u$ or $d$ with three-momentum $p$. The bare quark
mass $m$ is determined by fixing the pion mass $m_{\pi}$ through the dispersion relation
\begin{equation}\label{ee-NJL}
1-GJ_{pp}=0\;,
\end{equation}
with
\begin{equation}\label{eee-NJL}
J_{pp}=4(2I_{1}+q_{0}^{2}I_{2})\;,
\end{equation}
and $I_{1,2}$ are given by the integrals
\begin{equation}\label{I1}
I_1=\frac{N_{c}}{4\pi^{2}}\int_{0}^{\Lambda}\frac{p^{2}dp}{E}\;,
\end{equation}
\begin{equation}\label{I2}
I_2=-\frac{N_{c}}{4\pi^2}\int_0^\Lambda\frac{p^{2}dp}{E}\bigg(
\frac{1}{(E+q_0^2)^2-E^2}+\frac{1}{(E-q_0^2)^2-E^2}\bigg).
\end{equation}
Substituting the integrals (\ref{I1}) and (\ref{I2}) into the relation (\ref{eee-NJL})
leads to the following condition for the pion mass:
\begin{equation}\label{f-NJL}
\frac{2N_cG}{\pi^2}\int_{0}^{\Lambda}\frac{dp\,p^2}{E}\bigg(1-
\frac{q_0^2}{q_0^2-4E^2}\bigg)\bigg{|}_{q_0^2=m_\pi^2}=1,
\end{equation}
through the dispersion relation (\ref{ee-NJL}). From the above relations, it is clear
that the NJL model is nonrenormalizable and it is necessary to introduce a
three-momentum cutoff $\Lambda$ as
\begin{equation}\label{g-NJL}
\frac{1}{(2\pi)^3}\int d^3p=\frac{1}{2\pi^2}\int_0^{\infty}p^2
dp\hspace{.3cm}\underrightarrow{\mbox{NJL}}\hspace{.3cm}\frac{
1}{2\pi^2}\int_0^\infty\Theta[\Lambda-p]p^2dp,
\end{equation}
where $\Theta$ is the Heaviside step function which guaranties the fact that the
interaction described by the NJL Lagrangian (\ref{a-NJL}) would be valid just for the
momenta smaller than the cutoff $\Lambda$. Indeed, the interaction becomes weak for
the large momenta and then reproduces the asymptotic freedom in this regime.

Thus, the three parameters $\Lambda$, $G$ and $m$ related to NJL model should be
fixed by the physical quantities that are defined in relations (\ref{d-NJL}),
(\ref{e-NJL}) and (\ref{f-NJL}). We will use the data that are reported in Ref.
\cite{Hatsuda} through this paper (see Table 1).
\begin{table}
 \begin{center}
 \caption{\footnotesize Parameter set (the first and second rows) that are fixed by the
 physical quantities (the third and fourth rows). The constituent quark mass ends up
 with $M=325$ MeV with these data. The number of flavors and colors are taken to be
 $N_{f}=2$  and $N_{c}=3$.}

\begin{tabular}{|c|c|c|}
  \hline
  $\Lambda$ (GeV) & $G(GeV^{-2})$& $m$(MeV) \\ \hline
  0.651 & 10.08 & 5.5  \\  \hline
  $-\langle\bar{\psi}\psi\rangle^{1/3}$(MeV) &
  $f_{\pi}$(MeV)& $m_{\pi}$(MeV)  \\  \hline
  -251 & 92.3 & 139.3 \\
    \hline
\end{tabular}
\end{center}
\end{table}

As a final remark in this section, the parameters that are listed in Table 1 can reproduce
an effective mass about $M\approx325$ MeV which is corresponds approximately to the
one-third of the mass of the nucleons. Therefore, this set of parameters of the model can
explain the additional mass of the proton.

\section{Compact Momentum Spaces and Deformed Measure}
In this section, we consider the effects of a minimal length scale to the relation (\ref{g-NJL})
by which one can find QG corrections on the physical quantities (\ref{d-NJL}), (\ref{e-NJL}),
and (\ref{f-NJL}) in the NJL model.

Indeed, the QG phenomenological approaches to the issue of minimal length suggest different
deformations to the standard measure (\ref{g-NJL}). But, interestingly, there is a criterion
which is common between these alternative approaches: {\it Topology of the momentum space will
be compact in order to take into account an ultraviolet cutoff for the system under
consideration}, see Ref. \cite{C-PS} for more details. For instance, topology of the momentum
part of a two-dimensional polymerized phase space is a circle ${\mathbf S}^1$ rather than the
standard ${\mathbf R}$ topology \cite{CPR} and the radius of the circle is directly related to
the maximal momentum (or minimal length). While the four-momentum space in special relativity
has Minkowski geometry with ${\mathbf R}^4$ topology, it has de Sitter geometry with
${\mathbf R}\times{\mathbf S}^3$ topology in doubly special relativity scenarios where
${\mathbf R}$ is identified with the space of energy and ${\mathbf S}^3$ with the
three-momentum space. The compact ${\mathbf S}^3$ topology for the three-momentum space
induces an ultraviolet cutoff and indeed the radius of the three-sphere ${\mathbf S}^3$
determines the maximal three-momentum cutoff for the system under consideration \cite{AdS}.
Also, the three-momentum space of the non-relativistic Snyder model is constructed on
three-sphere ${\mathbf S}^3$ \cite{mignemi}. The constant curvature of maximally symmetric
spaces, which determines an invariant QG scale, makes them to be relevant candidates
for the three-momentum spaces in QG scenarios and, therefore, the three-sphere ${\mathbf S}^3$
is the most relevant choice since it preserves the rotational invariance. Thus, in the
semiclassical regime, transition to the theories which deal with minimal length scale is
possible just by replacing the standard measure of the three-momentum space by a measure that
is defined on a compact topology such as ${\mathbf S}^3$ as
\begin{equation}\label{t-measure2}
\int_{{\mathbf R}^3}d\mu(p)\hspace{.3cm}\underrightarrow{\mbox{QG}}
\hspace{.3cm}\int_{S^3}d\mu(p)=\int_{|p|<\frac{1}{\lambda}}\frac{
d^3p}{\sqrt{1-(\lambda p)^2}}\,,
\end{equation}
where $\lambda$ is the QG parameter with a dimension of length which signals the existence
of a maximal momentum (or minimal length) in this setup. It is widely believed that this
length scale will be of the order of the Planck length $\lambda=\lambda_0\,l_{_{\rm Pl}}$,
where $\lambda_0={\mathcal O}(1)$ is the dimensionless QG parameter which should be fixed
by experiments. The deformed measure (\ref{t-measure2}) is obtained in the context of the
polymer quantum mechanics \cite{poly-measure}, doubly special relativity theories
\cite{Girelli}, non-relativistic Snyder model \cite{mignemi} and also in the context of
the theory of stability of the Lie algebras \cite{mendes}.

The natural appearance of a three-momentum cutoff in deformed measure (\ref{t-measure2})
suggests the identification of $1/\lambda$ with three-momentum cutoff $\Lambda$ of
the standard NJL model. In this respect, the NJL model becomes naturally
ultraviolet-regularized. But, $1/\lambda$ should be of the order of the Planck energy
$E_{_{\rm Pl}}\sim\,10^{19}$ GeV (since we assume that it is a quantum gravitational
cutoff) while $\Lambda_{QCD} \approx100-200$ MeV. Therefore, $1/\lambda$ is no longer a
QG cutoff if one identifies it with $\Lambda$. We would like, however, to interpret it
as a QG parameter. Therefore, using (\ref{t-measure2}), we consider the following
QG-modified counterpart of (\ref{g-NJL}) as
\begin{equation}\label{measure}
\frac{1}{2\pi^2}\int_{0}^{\infty}p^2dp\hspace{.3cm}\underrightarrow{
\mbox{QG}}\hspace{.3cm}\frac{1}{2\pi^2}\int_{0}^{1/\lambda}
\frac{p^2dp}{\sqrt{1-(\lambda p)^2}}\hspace{.3cm}\underrightarrow{
\mbox{NJL}}\hspace{.3cm}\frac{1}{2\pi^2}\int_{0}^{1/\lambda}\frac{
\Theta[\Lambda-p]p^2dp}{\sqrt{1-(\lambda p)^2}}.
\end{equation}
The above relation allows us to find QG corrections to the quantities (\ref{d-NJL}),
(\ref{e-NJL}), and (\ref{f-NJL}) in NJL model and then bounding the quantum gravity
parameter $\lambda$.

\section{Quantum Gravity Corrections}

\subsection{Quark Condensate}
Using the QG-modified measure (\ref{measure}), the modification to the quark condensate
(\ref{d-NJL}) would be
\begin{equation}\label{d-QG}
\langle\bar{u}u\rangle=\langle\bar{d}d\rangle=-\frac{N_cM}{\pi^2}
\int_{0}^{\frac{p_{_{\rm Pl}}}{\lambda_0}}\frac{dp\,p^2\Theta[
\Lambda-p]}{E\sqrt{1-\lambda_0^2(p/p_{_{\rm Pl}})^2}},
\end{equation}
where we have substituted $\lambda=\lambda_0\,p_{_{\rm Pl}}^{-1}$. At the Planck scale,
the factor $\lambda_0(p/p_{_{\rm Pl}})\sim\,1$ and one then cannot expand the denominator
in terms of the QG parameter $\lambda$. However, the existence of the step function
guaranties that the momenta are bounded as $|p|\leq\Lambda$ and we have $\lambda_0(p/
p_{_{\rm Pl}})\sim\,10^{-20}\ll1$ and we then could always expand the denominator in this
setup as
\begin{equation}\label{b-QG}
\langle\bar{u}u\rangle=\langle\bar{u}u\rangle_0+\Delta\big(\langle\bar{u}u
\rangle\big)+{\mathcal O}\Big(\Big(\frac{p}{p_{_{\rm Pl}}}\Big)^4\Big)\;\;,
\end{equation}
where $\langle\bar{u}u\rangle_0$ is given by (\ref{d-NJL}) that is the quark condensate
in the absence of QG effects and we have also defined
\begin{equation}\label{c-QG}
\Delta\big(\langle\bar{u}u\rangle\big)=-\lambda_0^2\frac{N_cM}{2\pi^2
}\int_{0}^{\Lambda}\frac{dp\,p^2}{E}\Big(\frac{p}{p_{_{\rm Pl}}}\Big)^2,
\end{equation}
which is the QG correction to the quark condensate. Using the data presented in Table 1
and the numerical solution for the standard quark condensate (\ref{c-QG}), we obtain
the estimation for the QG correction to the quark condensate as
\begin{equation}\label{dd-QG}
\bigg{|}\frac{\Delta\big(\langle\bar{u}u\rangle\big)}{\langle\bar{u}u
\rangle_0}\bigg{|}=7.84\times10^{-40}\lambda_0^2\;\;.
\end{equation}
This result could be interpreted in two ways \cite{B-GUP1,B-GUP2}. First considering
$\lambda_0=1$, which guaranties that the QG effects become significant just at the
Planck scale. So, the QG corrections are very small to be detected for the accessible
energy scales (of the order of $10^{-40}$). The second interpretation is more
interesting. We could obtain an upper bound on $\lambda_0$ to ensure the validity of
the NJL estimation for the quark condensate up to the precision of the experiment
under consideration. The empirical value derived from QCD sum rules for quark condensate
has a precision of the order of $\approx10^{-2}$ \cite{Ratti, Shifman}. Given that the QG
correction (\ref{dd-QG}) to the quark condensate (\ref{d-NJL}) would be smaller than the
precision, so we get the following upper bound on $\lambda_{0}$
\begin{equation}\label{ee-QG}
\lambda_{0}\lesssim\,8\times10^{17}.
\end{equation}

\subsection{Pion Decay Constant}
By means of the deformed measure (\ref{measure}), the QG deformation to the integral
equation of the pion decay constant (\ref{e-NJL}) will be
\begin{eqnarray}\label{e-QG}
f_{\pi}^2=\frac{N_cM^2}{2\pi^2}\int_{0}^{\frac{p_{_{\rm Pl}}}{\lambda_0}}
\frac{dp\,p^2\Theta[\Lambda-p]}{E^3\sqrt{1-\lambda_0^2(p/p_{_{\rm Pl}})^2}}
\nonumber\\=f_{0\pi}^2+\Delta\big(f_{\pi}^2\big)+{\mathcal O}\Big(\Big(
\frac{p}{p_{_{\rm Pl}}}\Big)^4\Big)\;\;,
\end{eqnarray}
where $f_{0\pi}^2$ is given by the relation (\ref{e-NJL}) and in the same manner of the
quark condensate, we have defined the QG correction
\begin{equation}\label{f-QG}
\Delta\big(f_{\pi}^2\big)=\lambda_0^2\frac{N_cM^2}{4\pi^2}\int_{0}^{\Lambda
}\frac{dp\,p^2}{E^3}\Big(\frac{p}{p_{_{\rm Pl}}}\Big)^2\;\;.
\end{equation}
Substituting from the data in Table 1, we get the estimation
\begin{equation}\label{g-QG}
\bigg{|}\frac{\Delta\big(f_{\pi}^2\big)}{f_{0\pi}^2}
\bigg{|}=1.92\times10^{-42}\lambda_0^2\;\;.
\end{equation}
Again, we could set $\lambda_0=1$ that clearly leads to the very small QG correction to
the pion decay constant of the order of $10^{-42}$. But, in the same manner for the
quark condensate, we could find an upper bound on the QG dimensionless parameter
$\lambda_0$ through the precision of the experiment. Indeed, pion decay constant $f_{\pi}$
may be obtained from the decay rate $\pi^{+}\rightarrow\mu^{+}\nu$ so that its empirical
value is reported with a precision of $3000$ ppm (part per million) \cite{Yndurain}.
Despite the fact that the QG correction (\ref{g-QG}) should be less than the current
accuracy of the precision of the pion decay constant measurement, we obtain the following
upper bound on $\lambda_0$
\begin{equation}\label{h-QG}
\lambda_{0}\lesssim\,2.1\times10^{18},
\end{equation}
which is weaker than that we have obtained for the case of the quark condensate in
relation (\ref{ee-QG}).

\subsection{Pion Mass}
Taking the QG effects by means of the deformed measure (\ref{measure}) into account, the
integral relation for the pion mass (\ref{f-NJL}) will be modified as
\begin{equation}\label{p-PM}
\frac{2N_cG}{\pi^2}\int_{0}^{\frac{1}{\lambda}}\frac{dp\,p^2
\Theta[\Lambda-p]}{E\sqrt{1-(\lambda p)^2}}\bigg(1-\frac{q^2
}{q_0^2-4E^2}\bigg)\bigg{|}_{q_0^2=m_\pi^2}=1.
\end{equation}
Since we know that the QG effects are very small, we consider the ansatz $q=q_0+\lambda\,
\delta{q}=q_0(1+\lambda^2(\delta{q}/q_0))$, where $q_0$ solves the nondeformed relation
(\ref{f-NJL}) and $\lambda^2(\delta{q}/q_0)$ is the dimensionless quantity which signals
a very small QG effect. Substituting this ansatz into the relation (\ref{p-PM}) and then
using the nondeformed relation (\ref{f-NJL}), it is easy to show that to first order of
approximation ${\mathcal O}(\lambda^2)$, we have
\begin{eqnarray}\label{approx-pm}
\delta{q}=\frac{1}{16}\bigg\{2\Lambda\sqrt{\Lambda^2+M^2}(-3M^2+2
\Lambda^2+q_0^2)\hspace{9cm}\nonumber\\-(6M^4-6M^2q_0^2+q_0^4)\ln\Big[
\frac{M}{\Lambda+\sqrt{\Lambda^2+M^2}}\Big]+q_0(4M^2-q_0^2)^{3/2}
\tan^{-1}\Big[\frac{\Lambda\,q_0}{\sqrt{(\Lambda^2+M^2)(4M^2-q_0^2)}}
\Big]\bigg\}\nonumber\\{\times}\bigg\{\frac{2q_0\Lambda\sqrt{\Lambda^2
+M^2}}{4\Lambda^2+4M^2-q_0^2}+q_0\ln\Big[\frac{M}{\Lambda+\sqrt{
\Lambda^2+M^2}}\Big]+\frac{2M^2}{\sqrt{4M^2-q_0^2}}\tan^{-1}\Big[\frac{
\Lambda\,q_0}{\sqrt{(\Lambda^2+M^2)(4M^2-q_0^2)}}\Big]\bigg\}^{-1}.
\end{eqnarray}
The precision of reported empirical value of the pion mass is corresponding to the $2.5$ ppm
\cite{Data Group}. Using the numerical value of $q_0=m_\pi$ and also the data in Table
1, we obtain the numerical estimation
\begin{equation}\label{pmass-QG}
\lambda^2\bigg{|}\frac{\delta{q}}{q_0}\bigg{|}=8.69\times10^{-39
}\lambda_0^2\;\;,
\end{equation}
for the QG correction to the pion mass in this setup. As we have stated previously, while
this correction is very small to be measured, it could be also smaller than precision of
the measurement. Taking this fact into account, we obtain an upper bound for $\lambda_0$
as
\begin{equation}\label{pmass-QG2}
\lambda_{0}\lesssim\,1.4\times10^{16}\,,
\end{equation}
which is stronger than those obtained for the two pervious cases: Quark condensate and
pion decay constant in relations (\ref{ee-QG}) and (\ref{h-QG}) respectively.

\section{Summary and Conclusions}
It is widely believed that quantum gravity (QG) effects would become significant at the
Planck scale, where the energy scale of the system is comparable with Planck energy
$E_{_{\rm Pl}}\sim10^{19}$ GeV. Although full quantum theory of gravity has not been
formulated yet, QG candidates such as string theory and loop quantum gravity suggest the
existence of a minimal length scale. Taking a minimal length into account immediately leads
to the deformation of the algebraic structure of the quantum mechanics. In this respect,
effective models such as the generalized uncertainty principle, polymer quantum mechanics
and noncommutative quantum mechanics have been suggested which support the existence of a
minimum length as an ultraviolet cutoff for the system under consideration. The doubly
special relativity theories are also investigated which take into account an invariant
observer-independent length scale in special relativity. It is natural to expect that
the minimum length scale being of the order of the Planck length as $\lambda=\lambda_0
l_{_{\rm Pl}}$, where $l_{_{\rm Pl}}\sim\,10^{-35}$ m, is the Planck length and
$\lambda_0={\mathcal O}(1)$ is a dimensionless parameter which should be fixed by
experiment. However, a definite value for $\lambda_0$ should be determined through
very high energy regime experiments. Nevertheless, one can obtain upper bounds on
$\lambda_0$ (or equivalently on $\lambda$) through the accuracy of the measurement in
some well-known low energy experiments. For instance, QG parameter is constrained as
$\lesssim\,10^{17}$ in noncommutative geometry framework \cite{B-NC}. The upper bound
$\lesssim\, 10^{11}$ is obtained in the generalized uncertainty principle framework
\cite{B-GUP1} (see also Refs. \cite{B-GUP2} where the weaker bounds are obtained). In
the context of the polymer quantum mechanics, the upper bounds $\lesssim\,10^{27}$ and
$\lesssim\,10^{22}$ are obtained in Refs. \cite{B-POLY1} and \cite{B-POLY2} respectively.
Apart from the details of the these effective theories, the common feature between all
of them is the deformation of the momentum space such that it defines on the compact
topology. In this paper, considering a compact momentum space with three-sphere
${\mathbf S}^3$ topology, which is suggested by the doubly special relativity theories,
Snyder model and noncommutative spaces, we have studied the effects of the minimal
length on the determination of the physical quantities in the $SU(2)$
Nambu-Jona-Lasinio model of QCD. In particular, we found QG corrections to the
three physical quantities including the chiral condensate, pion decay constant and
pion mass. QG corrections would be less than the current accuracy of the measurement and
we then obtained upper bound $\lambda_0\lesssim\,1.4\times10^{16}$ for the QG
dimensionless parameter in this setup. Stronger bounds can be obtained by improving
precision of the measurements in this setup in future.

{\bf Acknowledgement}\\
We would like to thank an anonymous referee for very insightful comments.

\end{document}